\documentclass[preprint,prc,aps,superscriptaddress,nofootinbib,showpacs]{revtex4}
\usepackage[dvips]{graphicx}
\usepackage{amssymb}

\def\beq {\begin{equation}}
\def\eeq {\end{equation}}
\def\bea {\begin{eqnarray}}
\def\eea {\end{eqnarray}}

\def\nn {\nonumber}
\def\vs {\vspace}

\def\lp {\left( }
\def\rp {\right) }
\def\lb {\left[ }
\def\rb {\right] }
\def\lc {\left\{ }
\def\rc {\right\} }

\def\rar {\rightarrow}
\def\lrar {\leftrightarrow}

\def\Rb {\bar{R}}

\def\sp {\!+\!}
\def\sm {\!-\!}

\def\so {{_{1/2}}}

\def\cO {{\cal{O}}}

\def\k {\kappa}

\def\g {\gamma}

\def\p {\pi}
\def\s {\sigma}

\begin{document}
\vs{1cm}
\title{Elastic $K \p$ amplitude: a simple model}

%%%%%%%%%%%%%%%%%%%%%%%%%%%%%%%%%%%%%%%%%%%%%%%%%%%%%%%%%%%%%%%%%%%%%%%%%%%%%
\author{P. C. Magalh\~aes}
\email[]{patricia@if.usp.br}
\author{M. R. Robilotta}
\email[]{robilotta@if.usp.br}

\affiliation{Instituto de F\'{\i}sica, Universidade de S\~{a}o Paulo,\\
C.P. 66318, 05315-970, S\~{a}o Paulo, SP, Brazil.}

%%%%%%%%%%%%%%%%%%%%%%%%%%%%%%%%%%%%%%%%%%%%%%%%%%%%%%%%%%%%%%%%%%%%%%%%%%%%%%
\date{\today}

\begin{abstract}
We present a  chiral model for the $J=0,\; I=1/2,$ elastic $K\p$ 
amplitude, suited to be employed in $D^+ \rar K^- \p^+ \p^+$ data analyses and 
valid between threshold and $1.5\;$GeV.
Although not as precise as other versions available in the literature,
it is rather simple and incorporates the essential physics in this energy 
domain. 
In the case of the $K$-matrix approximation, the model allows the 
pole structure of the $K\p$ amplitude to be understood by solving a 
quadratic equation in $s$.
We show that the solutions to this equation can be well approximated 
by polynomials of masses and coupling constants. 
This analytic structure allows a clear understanding why, depending 
on the values of one of the coupling constants, one may have one or two
physical poles. 
The model yields a pole, associated with the $\k$, 
at $\sqrt{s}= (0.75 - i\, 0.24)\;$GeV.

\end{abstract}

\pacs{13.20.Fc, 13.25.-k, 11.80.-m} 

\maketitle

% section 1 OOOOOOOOOOOOOOOOOOOOOOOOOOOOOOOOOOOOOOOOOOOOOOOOOOOOOOOOOOOOOOOOOO
\section{introduction}

Since the E791 experiment\cite{E791}, heavy-meson decays have 
systematically produced solid evidence in favour of low-energy 
scalar resonances\cite{E791K,FOCUS,CLEOc}.
The quality of these results renewed  interest in the problem 
and motivated an effort aimed at determining the positions of their
poles in the complex energy plane.   
In the case of $\p \p$ resonances, the pole position of the $\s$ can be 
determined from scattering data by means of the Roy equation, and the
rather precise value 
$\sqrt{s_\s} = (0.441_{-8}^{+16} - i\, 0.272_{-12.5}^{+9})\;$MeV
is available\cite{Lpole}.

The situation of scalar resonances in the $K\p$ system is much less certain,
since our knowledge of the $S$-wave $I=1/2$ amplitude is based on just two
experiments\cite{Estab,Aston}, 
which include phase shifts only up to $\sim 80^0$.
These data sets have been carefully analyzed in recent years, 
and determinations of both pole positions\cite{JOP} and 
scattering lengths\cite{BDM} became available.
Nevertheless, the need of more empirical knowledge is still urgent and, 
in principle, information from the decay $D^+ \rar K^- \p^+ \p^+$ could 
be useful in either constraining or complementing $K\p$ scattering data.

\begin{figure}[h] 
\includegraphics[width=1.0\columnwidth,angle=0]{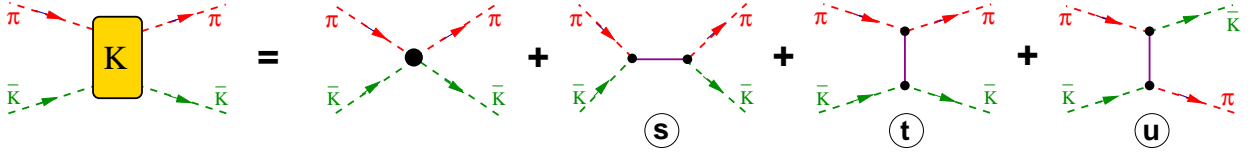}
\caption{(Color online) Tree-level $\pi K$ amplitude;
dashed and full lines represent respectively pseudoscalar mesons 
and  scalar resonances.}
\label{F1}
\end{figure}

Analyses of $D^+$ decays normally rely on trial functions written in terms of 
Breit-Wigner expressions, which are at odds with chiral symmetry.
As the symmetry is very important at low-energies, this kind of procedure
has been criticized and alternatives were proposed\cite{AR, Oller}.
In the work by Oller\cite{Oller}, data were refitted with the help of a 
chiral amplitude, with a remarkable decrease in $\chi^2$.
In a recent paper\cite{BMRZ}, we have discussed the main features of the
decay $D^+\to K^-\pi^+\pi^+$ at low energies, including  explicit 
descriptions of both the primary weak vertex and final state interactions, 
based on a unitarized $K\p$ amplitude.
This elastic amplitude was obtained by the iteration of a kernel by means
of a simplified Bethe-Salpeter equation\cite{OO97}.
The $(J,I)=(0,1/2)$ component of the kernel, which interests us here,
was derived from effective lagrangians and contains
a leading $\cO(q^2)$ contact term\cite{GL}, 
supplemented by an explicit resonance exchange\cite{EGPR},
corresponding to $\cO(q^4)$ corrections.
This model is illustrated in fig.\ref{F1}. 

%fig 1 ^^^^^^^^^^^^^^^^^^^^^^^^^^^^^^^^^^^^^^^^^^^^^^^^^^^^^^^^^^^^^^^^^^^^^^^
%^^^^^^^^^^^^^^^^^^^^^^^^^^^^^^^^^^^^^^^^^^^^^^^^^^^^^^^^^^^^^^^^^^^^^^^^^^^^^

Our ultimate goal is to produce a tool to be directly employed 
in analyses of raw data and therefore we try to avoid, as much as possible,
the use of long and involved expressions.
We thus neglect contributions from resonance exchanges in $t-$ and $u-$ 
channels, since they just give rise to small backgrounds\cite{BMRZ}.
In the present work we follow this approach and show that simplifications 
are possible in the chiral kernel which preserve its essential physical
content.

% section 2 OOOOOOOOOOOOOOOOOOOOOOOOOOOOOOOOOOOOOOOOOOOOOOOOOOOOOOOOOOOOOOOOO
\section{$K\p$ amplitude}

Our unitarized $(J, I)=(0,1/2)$ amplitude for the process 
$\p \, K \rar \p \, K$ has been discussed in detail in ref.\cite{BMRZ}
and here we just summarize its main features.
It is written as
\bea
&& T_\so(s) = \g^2(s)/D(s) \;,  
\nn\\
&& D(s)= [m_R^2 -s +\g^2(s) \,\Rb_\so(s)]   -i\,\lb\g^2(s) \,\frac{\rho(s)}{16 \p}\rb,
\label{kp.1}
\eea
where:
\\
- $s$ is the usual Mandelstam variable and 
$\rho(s)=\sqrt{1 - 2\, (M_K^2 \sp M_\p^2)/s+ (M_K^2 \sm M_\p^2)^2/s^2}\;$;
\\
- $m_R$ is the parameter present in the chiral lagrangian, 
called {\em nominal} resonance mass;
\\
- $\Rb_\so(s)$ is the function describing off-shell effects 
in the two-meson propagator, given by
\bea
\Rb_\so (s) &=& - \Re  \lb  L(s) - L(m_R^2) \rb /16\p^2 \;,
\nn\\[2mm]
 \Re L(s) &=& \rho(s) \log \lb (1 - \s)/(1 + \s)\rb 
- 2 \nn\\[2mm]&& +\, [(M_K^2 -M_\p^2)/s] \,\log( M_K/M_\p)] \;,
\nn\\[2mm]
 \s &=& \sqrt{|s \sm (M_K \sp M_\p)^2|/|s \sm (M_K \sm M_\p)^2 | } \;;
\label{kp.2}
\eea
- $\Rb_\so(m_R^2)=0\;$ by construction and therefore the phase shift is
$\pi/2$ at $s=m_R^2\;$;
\\
- $\g^2(s)$ is the function which incorporates chiral dynamics, 
given by 
\bea
\g^2(s) \!&=&\!   \lc (1 /F^2)\,
\lb \lp 1 -3\,\rho^2(s)/8 \rp s \right.\right.\nn\\&&-\left.\left. \lp M_\p^2 + M_K^2\rp \rb \,(m_R^2-s)\rc_L
\nn\\[2mm]
&&+\! \lc  (3/F^4) \,[c_d\, \lp s \sm M_\p^2 \sm M_K^2 \rp 
 \right.\nn\\&& + \left. c_m \, \lp 4\,M_K^2 \sp 5\,M_\p^2 \rp /6 ]^2 \rc_R \;;
\label{kp.3}
\eea
- the labels $L$ and $R$ in the curly brackets denote contributions from 
the leading $\cO(q^2)$ contact term and the $\cO(q^4)$ resonance correction;
\\
- $F$ is the pseudoscalar decay constant and the parameters $c_d$ and $c_m$ 
are the resonance couplings defined in  ref.\cite{EGPR}.

This model for the $K\p$ amplitude depends on six parameters,
three of which are well known and given by
$M_\p=0.1396\;$GeV, $M_K=0.4937\;$Gev and $F=\sqrt{F_\p F_K}=0.103\;$GeV.
The other three, namely $m_R$, $c_d$ and $c_m$ need to be taken from 
experiment. 
In ref.\cite{EGPR} one finds $(|c_d|,|c_m|)= (0.032, 0.042)\;$ GeV,
obtained from the process $a_0 \rar \eta\,\pi$.
We adopt these values provisionally and, at the end, suggest our own choice.

%fig 2 ^^^^^^^^^^^^^^^^^^^^^^^^^^^^^^^^^^^^^^^^^^^^^^^^^^^^^^^^^^^^^^^^^^^^^
\begin{figure}[h]
\begin{center}
\hspace*{-25mm}
\includegraphics[width=.7 \columnwidth,angle=0]{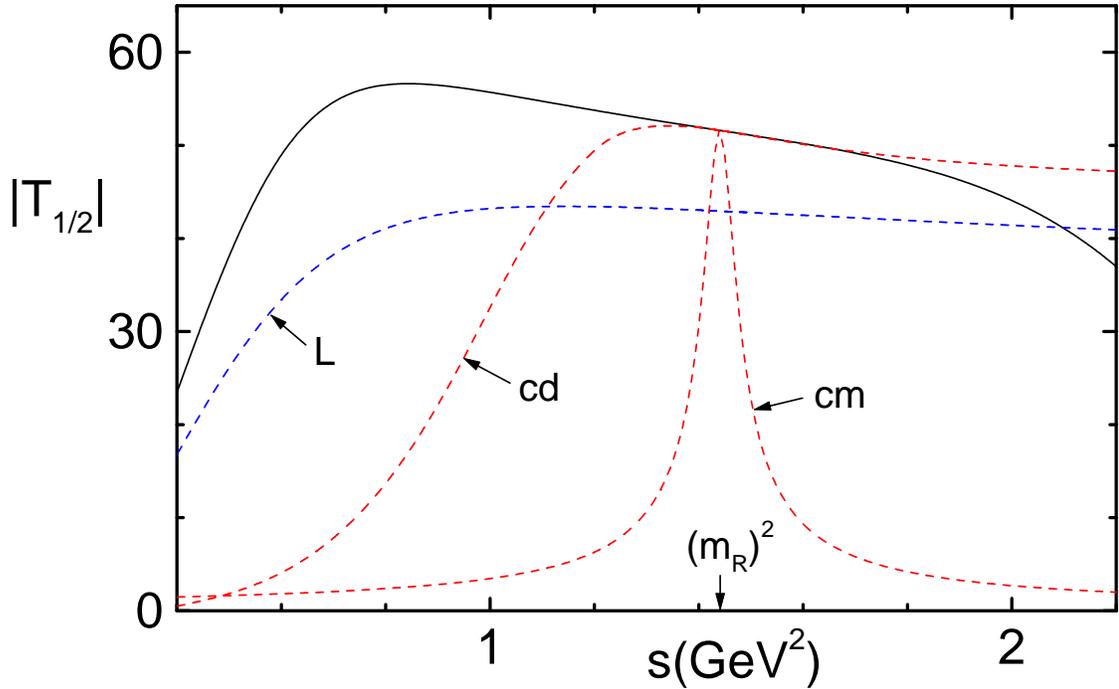}
\caption{(Color online) Modulus of the elastic amplitude:
the full curve represents eq.(\ref{kp.3}), the dashed curve $L$ 
is obtained by making $c_d=c_m=0$, 
and those labeled $c_d$ and $c_m$ are resonance contributions 
proportional to these parameters.} 
\label{F2} 
\end{center} 
\end{figure}
%^^^^^^^^^^^^^^^^^^^^^^^^^^^^^^^^^^^^^^^^^^^^^^^^^^^^^^^^^^^^^^^^^^^^^^^^^^^^

The influence of the various dynamical mechanisms over $|T_\so|$ is displayed
in Fig.\ref{F2}.
The full curve is based on eq.(\ref{kp.3}), whereas  
the dashed ones describe partial contributions:
$L$ corresponds to $c_d=c_m=0$ and those labeled $c_d$ and $c_m$ 
arise from just the terms proportional to these parameters.
The curve $c_m$ has a typical Breit-Wigner shape, but that labeled $c_d$
is much wider.
Nevertheless, both of them are constrained to be small at threshold, 
since the resonance corresponds to a $\cO(q^4)$ correction.
The leading $\cO(q^2)$ contact term clearly dominates at low-energies and,
as expected, full and resonance curves coincide at $s=m_R^2$,
since the resonance {\em nominal} mass is independent of coupling parameters.
By construction, $T_\so$ is purely imaginary at this point.

% section 3 OOOOOOOOOOOOOOOOOOOOOOOOOOOOOOOOOOOOOOOOOOOOOOOOOOOOOOOOOOOOOOOOOOO
\section{poles}

The poles of $T_\so$, determined by the condition $D(s)=0$, can be easily 
found out by numerical methods.
Before doing this, however, it is instructive to discuss the mathematical
structure of the problem, by means of a simplified version of the amplitude, 
in which:
\\
- the function $\Rb_\so$ is neglected, which corresponds to the 
$K-$matrix approximation;
\\
- the pion mass is neglected, which corresponds to the $SU(2)$ limit.
\\
The poles can then be found by solving the complex quartic equation
\bea
&& \lb 5/8 - \frac{3c_d^2}{F^2}\rb s^4
+[-(5m_R^2+7M_K^2)/8  \nn\\&&+\frac{c_d}{F^2}(9c_d - 4 c_m)M_K^2+ i\,16\p F^2]s^3
+ [(7m_R^2-M_K^2)\frac{M_K^2}{8}  \nn\\&&- ( c_d-2c_m/3)(9c_d-2c_m)\frac{M_K^4}{F^2} 
- i\,16\p F^2 m_R^2]s^2 
\nn\\
&& + [(m_R^2+3M_K^2)/8+3(c_d-2c_m/3)^2 M_K^2/F^2]M_K^4\, s \nn\\
&&-3 m_R^2 M_K^6/8 = 0 \;.
\label{kp.4}
\eea

Around physical poles, the quantity $M_K^2/|s|^2$ is small and one 
obtains the quadratic equation
\bea
&& A\;s^2 + B\; s + C =0  \;,
\label{kp.5}\\
&& A = [5/8 - 3c_d^2/F^2]\;,
\nn\\
&& B = [-(5m_R^2+7M_K^2)/8 +c_d(9c_d - 4 c_m)\frac{M_K^2}{F^2} + i\,16\p F^2] \;,
\nn\\
&& C=  [7M_K^2/8 - i\,16\p F^2]m_R^2\;.
\nn
\eea
The parameter $A$ plays an important role in this problem and receives 
contributions from both the leading contact term and the resonance.
In particular, the condition $A=0$ yields $c_d=F\sqrt{5/24}=0.047\,$GeV,
which is jut 50\% larger than accepted empirical values.
In the sequence, we discuss the behavior of the solutions of this 
equation as functions of $A$, in the interval 
$5/8 \leq A \leq 0$. 
Exact solutions are available in its two extremes:
\\
- $A=5/8 \rar c_d=0:$ in this case, the resonance $R$ corresponds to a 
bound state in the real axis, which does not couple with the $K\p$ system, 
and one has 
\beq
s_+(5/8) = m_R^2 \;\;\;\;\; \mathrm{and} \;\;\;\;\; 
s_-(5/8) = \lb 7 M_K^2 / 5 - i \; 128 \p \, F^2  /5 \rb \;;
\label{kp.6}
\eeq
- $A=0:$ the dynamical equation is no longer quadratic and its single 
solution reads
\beq
s_-(0) = \frac{[7M_K^2/5 - i\;128 \p \, F^2/5]}
{1 + \lb\frac{7M_K^2}{5} - 8 c_d(9c_d - 4 c_m)\frac{M_K^2}{F^2} - i\; 128 \p \frac{F^2}{5}\rb/m_R^2}\;.
\label{kp.7}
\eeq

We note that these two solutions already show 
a prominent feature of the problem, namely the stability of the
solution $s_-(A)$ in the whole interval considered.
As $m_R^2$ is a large parameter, $s_-(0)=s_-(5/8)+\cO(1/m^2_R)$.

In the general case, the solutions of eq.(\ref{kp.5}) will have the form
\beq
s_\pm = [-B \pm \sqrt{D}\;]/2A 
\;\;\;\;\lrar \;\;\;\;
D=B^2-4AC\;.
\label{kp.8}
\eeq
The square root prevents algebraic simplification of results.
However, at the point $A=-\Re B/2m_R^2\sim 5/16$, one has $\Im D=0$ and 
an approximate solution can be obtained for $\sqrt{D}\;$. 
By imposing a quadratic polynomial in $A$ to interpolate this function
at $A=5/8$, $-\Re B/2m_R^2$ and $0\;$, one finds 
\bea
 s_+&=& \frac{1}{A}\lc \lb \frac{5}{8}\, m_R^2 
- \frac{c_d}{F} \lp \frac{24 c_d}{5F}-\frac{4c_m}{F} \rp M_K^2
\right.\right.\nn\\&&- \left.\left. \frac{3c_d^2}{m_R^2 F^2} \lp 1- \frac{24c_d^2}{5F^2}\rp 
\lp \frac{128\p F^2}{5}\rp^2  \rb \right.
\nn\\[2mm]
&& \left. - i \; \frac{c_d}{F}\lb 3 \frac{c_d}{F}
- \lp \frac{3 c_d}{5F}-\frac{4 c_m}{F} \rp \frac{M_K^2}{m_R^2}
\right.\right.\nn\\&&- \left.\left. \frac{3 c_d}{F} \lp 1- \frac{24c_d^2}{5F^2}\rp 
\lp\frac{128\p F^2}{5 m_R^2}\rp^2\rb\frac{128\p F^2}{5}\rc,
\label{kp.9}\\[2mm]
 s_- &=& \lc \lb \frac{7}{5}\, M_K^2 
+ \frac{24 m_R^2 c_d^2}{5F^2}\lp \frac{128\p F^2}{5 m_R^2}\rp^2 \rb 
\right.\nn\\&&- \left. i\; \lb 1 - \frac{24 c_d^2}{5F^2}\lp \frac{128\p F^2}{5 m_R^2}\rp^2 \rb
\frac{128\p F^2}{5} \rc \;.
\label{kp.10}
\eea  

In tables \ref{T1} and \ref{T2} we display figures derived from 
eqs.(\ref{kp.4}), (\ref{kp.5}) and (\ref{kp.9}-\ref{kp.10}), 
for a wide sample of values for $c_d/F$. 
We recall that empirical estimates of this quantity lie in a narrow band 
around $c_d/F \sim 0.3$ and the condition $A=0$ corresponds to 
$c_d/F=0.456$.
As far as accuracy is concerned, one learns that predictions from
the quartic and quadratic equations deviate very little. 
The approximate algebraic solutions are less precise but, nevertheless, describe well
the qualitative features of the results in the whole range considered 
and are reasonably precise in the region of empirical interest.
The accuracy of eqs.(\ref{kp.9}-\ref{kp.10}) could be improved by keeping more terms
in series expansions, but this would just yield larger expressions,
without further conceptual gains.

%xxxxxxxxxxxxxxxxxxxxxxxxxxxxxxxxxxxxxxxxxxxxxxxxxxxxxxxxxxxxxxxxxxxxxxxxxxxxxx
\begin{table}[h]
\caption{Poles $\sqrt{s_+}$, in GeV, of the amplitude $T_\so$, for 
$m_R=1.2\,$GeV and $c_m=0.042\,$GeV.}
\begin{tabular} {|c|c|c|c|c|} \hline
 & quartic & quadratic & analytic & simplified \\ 
$\,c_d/F\,$ & eq.(\ref{kp.4}) & eq.(\ref{kp.5}) & eq.(\ref{kp.9}) & eq.(\ref{kp.11})
\\ \hline \hline
0   & $1.2000$  		  &$1.2000$ 			 & $1.2000$           & $1.2000$ \\ \hline
0.1 & $1.2351 -i\,0.0269$ & $1.2357 -i\,0.0272$ & $1.2394 -i\,0.0269$ & $1.2300 -i\,0.0175$\\ \hline
0.2 & $1.3135 -i\,0.0898$ & $1.3155 -i\,0.0918$ & $1.3257 -i\,0.0874$ & $1.3371 -i\,0.0758$\\ \hline
0.3 & $1.5346 -i\,0.2501$ & $1.5346	-i\,0.2500$ & $1.5513 -i\,0.2245$ & $1.6050 -i\,0.2022$\\ \hline
0.4 & $2.4694 -i\,0.6287$ & $2.4644 -i\,0.6327$ & $2.4566 -i\,0.6408$ & $2.5521 -i\,0.5534$\\ \hline
0.5 & $0.8660 +i\,2.7361$ & $0.8586 +i\,2.7414$ & $1.0882 +i\,2.8707$ & $0.9040 +i\,2.8315$\\ \hline

\end{tabular}
\label{T1}
\end{table} 
%xxxxxxxxxxxxxxxxxxxxxxxxxxxxxxxxxxxxxxxxxxxxxxxxxxxxxxxxxxxxxxxxxxxxxxxxxxxxxx

%xxxxxxxxxxxxxxxxxxxxxxxxxxxxxxxxxxxxxxxxxxxxxxxxxxxxxxxxxxxxxxxxxxxxxxxxxxxxxx
\begin{table}[h]
\caption{Poles $\sqrt{s_-}$, in GeV, of the amplitude $T_\so$, for 
$m_R=1.2\,$GeV and $c_m=0.042\,$GeV.}
\begin{tabular} {|c|c|c|c|} \hline
 & quartic & quadratic & analytic \\ 
$\,c_d/F\,$ & eq.(\ref{kp.4}) & eq.(\ref{kp.5}) & eq.(\ref{kp.10}) 
\\ \hline \hline
0   & $0.7908 -i\,0.5294$ & $0.7908 -i\,0.5294$ & $0.7908 -i\,0.5294$  \\ \hline
0.1 & $0.8064 -i\,0.5109$ & $0.8014 -i\,0.5173$ & $0.8001 -i\,0.5242$  \\ \hline
0.2 & $0.8458 -i\,0.4819$ & $0.8390 -i\,0.4879$ & $0.8206 -i\,0.4848$ \\ \hline
0.3 & $0.8904 -i\,0.4085$ & $0.8908	-i\,0.4126$ & $0.8586 -i\,0.4215$ \\ \hline
0.4 & $0.8756 -i\,0.3228$ & $0.8837 -i\,0.3165$ & $0.9189 -i\,0.3391$ \\ \hline
0.5 & $0.8440 -i\,0.2736$ & $0.8576 -i\,0.2575$ & $1.0041 -i\,0.2459$ \\ \hline

\end{tabular}
\label{T2}
\end{table} 
%xxxxxxxxxxxxxxxxxxxxxxxxxxxxxxxxxxxxxxxxxxxxxxxxxxxxxxxxxxxxxxxxxxxxxxxxxxxxxx

Conceptually, the behaviours of $\sqrt{s_+}$ and $\sqrt{s_-}$ 
shown in the tables are strikingly different.
The latter is a rather slow-varying function, whereas the former changes 
rapidly and even has the sign of the imaginary part reversed in the last row.
In order to understand this behavior of $\sqrt{s_+}$, we keep just the 
leading term in eq.(\ref{kp.9}) and find
\beq
s_+ = \frac{1}{A}\lc \frac{5}{8}\, m_R^2 
- i \; \frac{384\p}{5}\,c_d^2 \rc \;.
\label{kp.11}
\eeq
In spite of its simplicity, this result yields quite reasonable predictions, 
as indicated in the last column of table \ref{T1}.
Moreover, it shows clearly that the major features of $s_+$
are determined by the factor $A$ in the denominator.
In particular, it explains the change in the sign of the imaginary part 
of $\sqrt{s_+}$ observed in the bottom line of the table,
since $A$ vanishes at $c_d/F=0.456$.
As this factor does not occur in $s_-$, table \ref{T2} is much more monotonic. 

The following scenario is supported by 
eqs.(\ref{kp.9}-\ref{kp.10}):
\\
- in case the resonance $R$ is absent, one has just the pole 
$\sqrt{s_-}$, which is due to the leading contact interaction;  
\\
- if the resonance is present, but its couplings to mesons are not turned 
on ($c_d=c_m=0$), one has the pole $\sqrt{s_-}$ and a bound state 
in the real axis at $s=m^2_R$\cite{Eef2};
\\
- when the parameters $c_d$ and $c_m$ are turned on and the resonance $R$ couples
to $\sqrt{s_-}$,  both the mass and width of $\sqrt{s_+}$ {\em increase} 
monotonically, driven by the factor $A$ in the denominator;
\\
- as a consequence, for realistic values of $c_d$, one has necessarily
$\Re \sqrt{s_+}>m_R$ and the pole on the complex plane has to stand
on the right of the point at which the phase shift passes through $\p/2$;
\\
- the pole $\sqrt{s_+}$ blows up at the critical value $c_d/F=\sqrt{5/24}$
and, beyond this point, just $\sqrt{s_-}$ is present;

Although the couplings of the resonance $R$ 
to mesons is implemented by both $c_d$ and $c_m$, the former is much more
important than the latter, for it is incorporated into the parameter $A$.
The term proportional to $c_m$ is less relevant, because
it contains meson masses and vanishes  in the chiral limit.
For instance, if one chooses $c_d/F=0.3$ and sets $c_m=0$, 
the solutions of the quadratic equation become 
$\sqrt{s_+}= 1.3631 - i\,0.1973\;$GeV and 
$\sqrt{s_-}= 0.9971 - i\,0.4835\;$GeV.
This means that gentle variations of $c_m$ around empirical averages
would have little influence over pole positions. 

\begin{figure}[h]
\begin{center}
%\hspace*{-35mm}
\includegraphics[width=1 \columnwidth,angle=0]{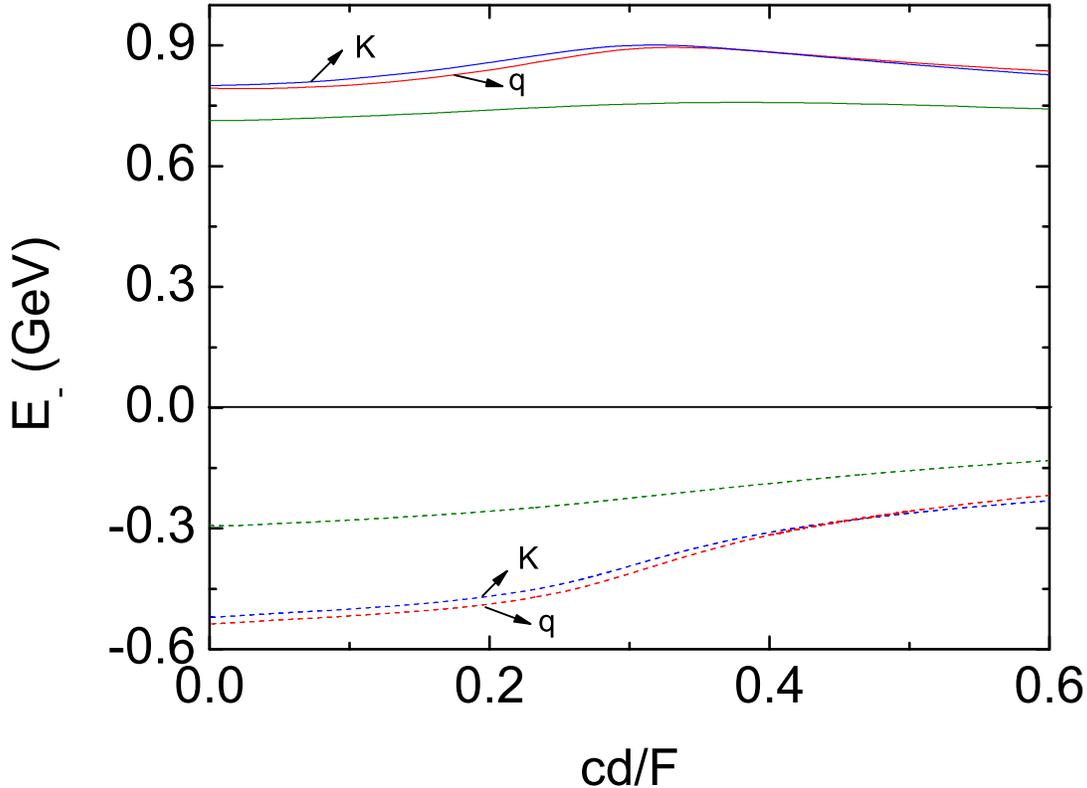}
\caption{(Color online) Real (full lines) and imaginary (dashed lines)
components of the function $E_-= \sqrt{s_-}$; the full curve comes for eq(\ref{kp.1}), whereas those labelled $K$ and $q$ correspond to the K-matrix and quadratic approximation.} 
\label{F3} 
\end{center} 
\end{figure}

\begin{figure}[h]
\begin{center}
%\hspace*{-35mm}
\includegraphics[width=1 \columnwidth,angle=0]{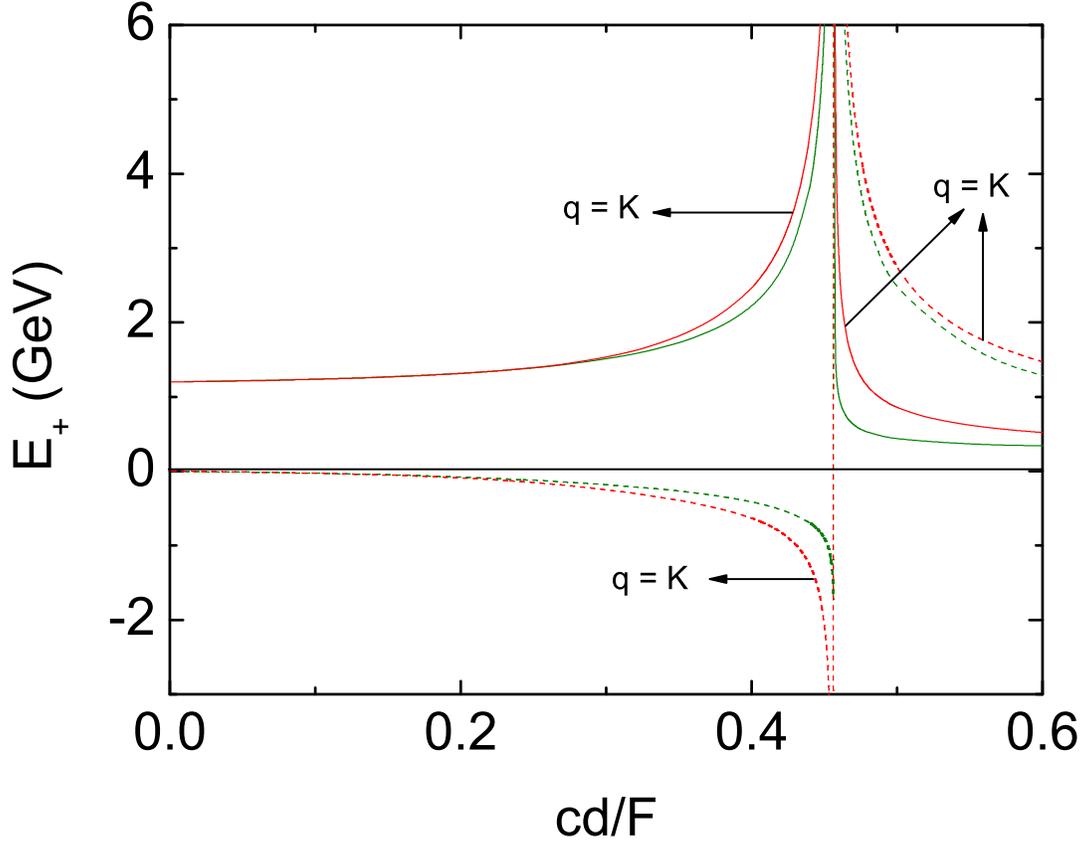}
\caption{(Color online) Real (full lines) and imaginary (dashed lines)
components of the function $E_+= \sqrt{s_+}$; the full curve comes for eq(\ref{kp.1}), whereas those labelled $K$ and $q$ correspond to the K-matrix and quadratic approximation.} 
\label{F4} 
\end{center} 
\end{figure}

In figs. \ref{F3} and \ref{F4}, predictions from the full model for $E=\sqrt{s}$, given by eq.(\ref{kp.1}), are compared with results from K-matrix and the quadratic 
approximations.
Quartic and quadratic approximations cannot be distinguished 
visually and the former is not shown.
Inspecting these figures, one learns that the inclusion of the pion mass is not numerically important, but off-shell effects in the two-meson propagator do influence the 
positions of the poles and tend to decrease both masses and widths.
Nevertheless, it does not alter the qualitative features 
of the scenario discussed above.

%fig 3 ^^^^^^^^^^^^^^^^^^^^^^^^^^^^^^^^^^^^^^^^^^^^^^^^^^^^^^^^^^^^^^^^^^^^^
%^^^^^^^^^^^^^^^^^^^^^^^^^^^^^^^^^^^^^^^^^^^^^^^^^^^^^^^^^^^^^^^^^^^^^^^^^^^^
%fig 4 ^^^^^^^^^^^^^^^^^^^^^^^^^^^^^^^^^^^^^^^^^^^^^^^^^^^^^^^^^^^^^^^^^^^^^
%^^^^^^^^^^^^^^^^^^^^^^^^^^^^^^^^^^^^^^^^^^^^^^^^^^^^^^^^^^^^^^^^^^^^^^^^^^^^

Our results suggest that the pole $\sqrt{s_+}$ can be 
identified with the $K_0^*(1430)$. 
The fitting of both the mass and width of this state supplies two constraints
for the resonance parameters.
As $c_m$ has little influence over numerical results, we fix it at the value 
$c_m=0.042\;$GeV and choose $m_R$ and $c_d$ so that 
$\sqrt{s_+}= [(1.414\pm 0.006) -i\,(0.145 \pm 0.010)] \;$GeV \cite{PDG}.
We then find $c_d/F = 0.2705 \pm 0.0078 \;$, $m_R= 1.1865 \pm 0.079 \;$GeV and
$\sqrt{s_-}= (0.7505 \pm 0.0010) - i\, (0.2363 \pm 0.0023)\;$GeV.
It is interesting to compare our results with those from
the much more complete analysis by Jamin, Oller and Pich\cite{JOP}.
Although there are no data in the $(J, I) = (0, 1/2)$ channel for phase shifts 
around $\p/2$, parametrizations produced in their paper suggest
that $m_R \sim 1.32\;$GeV and our value misses this point by 10\%.
On the other hand, by fixing the mass of the $K_0^*$ and imposing  
$m_R = 1.32\;$GeV, we find $c_d/F = 0.1640 $, 
$\sqrt{s_-}= 0.7419 - i\, 0.2541 \;$GeV, and 
$\Gamma_{K_0^*} = 0.1035\;$GeV, which is too small.
In both cases, the values of $c_d/F$ are below those of ref.\cite{EGPR}.
The difficulty of fitting simultaneously $m_R$ and $\Gamma_{K_0^*}$ may be
associated with the fact that we did not include inelasticities
present in the region $s>1\;$GeV.  

For the sake of completeness, in table \ref{T3}, we compare our result, $\sqrt{s_-}= 0.751 - i\,0.236 \,\,GeV$, with the positions of the $\k$ pole obtained in previous works.

\begin{center}

\begin{table}[h]
\caption{Pole position of the $\k$, in GeV.}
\begin{tabular} {|c|c|c|}
\hline
year  & ref. & $\k$  \\
\hline
$1986$ & \cite{Eef86} & $0.727 - i\, 0.263$ \\ \hline
$1997$ & \cite{Ishida} & $0.905^{65}_{30} -i\,0.222^{115}_{55}$ \\ \hline
$1998$ & \cite{Black} & $0.911 - i \,0.158$ \\ \hline
$1999$ & \cite{OO99} & $0.779 - i \, 0.330$ \\ \hline
$2000$ & \cite{JOP} & $0.708 -i \, 0.305$ \\ \hline
$2002$ & \cite{E791K}& $0.721\pm19\pm 44 -i\,0.292\pm 21 \pm 44$ \\ \hline
$2003$ & \cite{Bugg} & $0.750^{30}_{55}\, - i \, 0.342 \pm 60 $ \\ \hline 
$2004$ & \cite{Pelaez}& $0.750\pm 18 -i\,0.226 \pm 11$  \\\hline
$2006$ & \cite{Zhou} & $ 0.694 \pm 53 -i\,0.303 \pm 30$ \\ \hline
$2006$ & \cite{Moussallan} & $0.658\pm 13 -i\,0.279\pm 12$  \\ \hline
$2006$ & \cite{BES2006} & $0.841\pm 23^{64}_{55} -i\,0.306\pm26^{27}_{44}$ \\ \hline
$2008$ & \cite{Rupp} & $0.772 -i \, 0.281 $ \\ \hline 
%%$2008$ & nois!!!& $78? -i3?? $    \\ \hline 
\end{tabular}
\label{T3}
\end{table}
\end{center}

%OOO summary 000OOOOOOOOOOOOOOOOOOOOOOOOOOOOOOOOOOOOOOOOOOOOOOOOOOOOOOOOOOOOOO
\section{summary}
 
In this work we have discussed a model for the $K\p$ amplitude in the 
$(J,I)=(0,1/2)$ channel, suited to energies up to $1.5\;$GeV.
It is aimed at being used in data analyses of processes such as
$D^+ \rar K^- \p^+ \p^+$ and given by
\bea
T_\so \!&=&\! \frac{\g^2}
{(m_R^2 -s)  -i\,\g^2 \,\rho/(16 \p)}\;,
\label{kp.f1}\\[3mm]
&&\g^2 \!=\! \frac{(s-M_K^2)}{F^2\,s} 
\lc \lp \frac{5\,s}{8} + \frac{3\,M_K^2}{8} \rp \lp m_R^2-s \rp
\nn \right.\\ &&+\left. s\,\frac{c_d}{F} 
\lb \frac{3\,c_d}{F} \lp s-M_K^2 \rp + \frac{4\,c_m}{F}M_K^2 \rb \rc \;.
\nn
\eea
It represents a compromise between simplicity and the essential 
phenomena of this channel.
The factor $(s-M_K^2)$ in $\g^2$ is the Adler zero, which arises naturally
in the framework of chiral symmetry.
It allows automatically for one complex pole, identified with the $\k$.
Depending on the values obtained for the free parameters 
$m_R$, $c_d$ and $c_m$, another pole can be present,
associated with the $K_0^*$.

\begin{acknowledgments}
PCM would like to thank Eef van Beveren and George Rupp for a kind invitation to visit their group in Portugal, for financial support, for friendly hospitality and for fruitful discussions. 

This work is supported by FAPESP (Brazilian Agency).
\end{acknowledgments}

%PCM would like to thank Eef van Beveren and George Rupp for the friendly hospitality and for  inviting and supporting PCM in Portugal, where fruitful discussion (has helped to improve) ou (has improved, talvez melhor) this work.

% REFERENCES OOOOOOOOOOOOOOOOOOOOOOOOOOOOOOOOOOOOOOOOOOOOOOOOOOOOOOOOOOOOOOOOO

\end{document}